# WBVR Photometry of the CBS HZ Her/Her X-1

© 2008 A.N. Sazonov
Sternberg Astronomical Institute, Universitetskii pr. 13, Moscow, 119991 Russia

We report four-color WBVR photoelectric photometry of the close binary system (CBS) HZ HER/HER X-1 made in 1986-1994. Our photometry usually covered at least two $35^d$ precession periods during each of the 1986-1990, 1992, and 1994 observing seasons. The accuracy and duration of our 10-year long photoelectric observations of the star with an x-ray source made it possible to study its long-term behavior. We refine some of the "fine" photometric effects on the light curves of the CBS and try to interpret them in terms of the model of mass transfer from the optical component of the CBS to the accretion disk (AD) of the neutron star (NS).

The resulting model of the system satisfactorily explains the irregularities of the gas flow, the “hot spot”, as well as individual “splashes” moving in their own Keplerian trajectories about the external regions of the AD of the NS in Her X-1. All the above effects show up in our photometry.

We report a series of light curves for the time period covered by our observations, as well as the color-color diagrams, which reflect the underlying physics of fine photometric effects. We use our WBVR data to qualitatively estimate the high-temperature radiation of the AD at different phases of the 35-day cycle. We also qualitatively interpret the results in terms of the model of the accretion disk of the NS precessing in the direction of the orbital motion of the system.

During the observing seasons in question we recorded conspicuous albeit short surges of radiation (with the duration of up to 10 minutes) at the primary minimum near orbital phases ± 0.015-0.025 and with the R-,V-,B-,and W-band amplitudes equal to $0^m.10$, $0^m.15$, $0^m.20$, and $0^m.25$, respectively. The light curve exhibits characteristic bends at these orbital phases, where the rate of magnitude variation changes rather sharply (Cherepashchuk, 1975). The times of such bends correlate with the appearance of "dips" on the x-ray light curve, which are most likely associated with a certain mode of mass transfer from the optical component of HZ Her onto the NS (Crossa, 1980). These characteristic bends and the inflection points at the same orbital phases at Min I appear almost at all precession phases of the 35-day cycle during the 11-year period covered by our observations of HZ HER/HER X-1.

Of special interest are observations of the system made during the primary minimum near orbital phases from $\varphi=0.97$ to $\varphi=0.03$ (here we consider the configuration of the Roche lobe with libration point $L_2$ located at the outer edge of the AD of the NS). An analysis of all these photometric features yields a detailed light curve of the system near Min I, and, in particular, “sharp” minima are observed during some seasons. These minima must be caused by the eclipses of the gaseous condensations at libration point $L_2$ heated by the x-ray flux of the companion (KIlyachkov et al., 1978, 1980, 1994; Bochkarev et al., 1987; Kippenhahn et al., 1979).

In the case of the Roche-lobe configuration with triangular libration points $L_4$ and $L_5$ the accretion features heated by the x-ray flux may be located at the orbital phases of $\varphi=0^p.166$ and $\varphi=0^p.833$, respectively. The 1987, 1989, and 1991-1994 light curves exhibit a certain photometric feature resembling a gaseous condensation or a “blob” in the orbital phase intervals $\varphi=0.158$-$0.168$ and $\varphi=0.830$-$0.853$ (Kippenhahn et al., 1979; Crosa, Boynton, 1980). The x-ray flux shows deep flickering on a time scale of several minutes during the "dips" within the ±0.15 orbital-phase interval (Vrlitek and Halpern, 1985). This flickering may be due to whatever gaseous features located in the system.

The light curves in all photometric bands have a flat plateau at Min I. The behavior of the light curves is somewhat chaotic near Min II, especially in the UV, and it should be interpreted as a manifestation of the optically thick and twisted accretion disk. It results in observable shadow effects, physical variability of accretion features, and manifestations of gaseous flows in the system, which, in turn, are initiated by various modes of mass transfer from the optical component of the CBS. During the 1988, 1990, and 1992 observing seasons minimum W-B color index values were recorded (W-B $\approx 0$ at the precession phase of $\psi=0.86$), which correspond to the spectral type of the optical component at Min I and to the deficit of UV flux. During the1989 observations the "hot" spot in the low state of the x-ray flux had minimum size, but the temperature of the spot was very high (according to the (W-B) color diagram for the year in question). The V-band luminosity of the accretion disk in the “on” state was equal to that of an A7-type subgiant, whereas the W-band luminosity far exceeds the W-band luminosity of an A7-type giant. The size of the spot in the "off" state did not exceed 0.4-0.5 radii of an A7-type star and its temperature varied in the $(23\text{-}24)\times 10^3$ K interval.

Near Min I the object brightened even before it emerged from the x-ray eclipse of the AD of the NS. This appears to be due to the “blobs” appearing, which are carried by rotational motion with the period of $P_{circ}=15\pm 3^h$ at the outer edge of the accretion disk (Karitskaya et al., 1986; Sazonov & Shakura, 1986), and to the fact that the cooling relaxation time is on the order of one to two rotations about the center of mass of the system. The scatter of the (V-R) color index exceeds that of the (B-V) index during all observing years.

The light curve shows several points in the “on” state near the x-ray eclipse. These points suggest that the asymmetry has the same sign and became somewhat stronger during the high state. An analysis of the (W-B), (B-V), and (V-R) color indices suggests that the hot spot was possibly located on the trajectory of the motion of the accretion feature across the limb of the optical component of the system. During most of the time of the $35^d$ cycle the trailing side of the accretion feature was several times brighter than the leading side. Kilyachkov et al. (1994) came to a similar conclusion.

We thus conclude that:
- The accretion rate onto the neutron star somewhat increased in 1986;

- Short surges with the amplitudes of $0^m.10$, $0^m.15$, $0^m.25$, $0^m.35$ in the R-, V-, B-, and W-band light curves, respectively, were observed in 1987 near orbital phases 0.015, and this behavior must be due to the projection of some of the hot blobs of the accretion feature onto the limb of the hot component of the CBS;

- Min II showed somewhat chaotic behavior during the 1988,1990 observing season, especially in the UV, and it should be interpreted as a manifestation in space of an optically thick and twisted accretion disk. The total V-band light amplitude near Min II reached about ~ $0^m.55$;

- The position of the “hot spot ” near Min II changed in 1992 with respect to the center of the optical component, however, its size did not change compared to 1990;